\documentclass[showpacs,preprintnumbers]{revtex4}
\usepackage{amssymb}
\usepackage{graphicx}
\usepackage[tbtags]{amsmath}
\usepackage[dvips]{color}

\begin{document}

\title{Distinguishing quantum from classical oscillations in a driven phase
qubit}

\begin{abstract}
Rabi oscillations are coherent transitions in a quantum two-level system
under the influence of a resonant drive, with a much lower frequency
dependent on the perturbation amplitude. These serve as one of the
signatures of quantum coherent evolution in mesoscopic systems. It was shown
recently [N. Gr\o nbech-Jensen and M. Cirillo, Phys. Rev. Lett. \textbf{95},
067001 (2005)] that in phase qubits (current-biased Josephson junctions)
this effect can be mimicked by classical oscillations arising due to the
anharmonicity of the effective potential. Nevertheless, we find \emph{%
qualitative} differences between the classical and quantum effect. First,
while the quantum Rabi oscillations can be produced by the subharmonics of
the resonant frequency $\omega _{10}$ (multiphoton processes), the classical
effect also exists when the system is excited at the overtones, $n\omega
_{10}$. Second, the shape of the resonance is, in the classical case,
characteristically asymmetric; while quantum resonances are described by
symmetric Lorentzians. Third, the anharmonicity of the potential results in
the negative shift of the resonant frequency in the classical case, in
contrast to the positive Bloch-Siegert shift in the quantum case. We show
that in the relevant range of parameters these features allow to confidently
distinguish the bona fide Rabi oscillations from their classical Doppelg\"{a}%
nger.
\end{abstract}

\date{\today }
\author{S N Shevchenko$^{1}$, A N Omelyanchouk$^{1}$, A M Zagoskin$^{2,3,4}$%
, S Savel'ev$^{2,4}$ and Franco Nori$^{2,5}$}
\affiliation{$^{1}$B.Verkin Instiute for Low Temperature Physics and Engineering, 47
Lenin Ave., 61103, Kharkov, Ukraine\\
$^{2}$Advanced Science Institute, RIKEN, Wako-shi, Saitama 351-0198, Japan\\
$^{3}$Department of Physics and Astronomy, The University of British
Columbia, Vancouver, B.C., V6T 1Z1 Canada\\
$^{4}$Department of Physics, Loughborough University, Loughborough LE11 3TU,
United Kingdom\\
$^{5}$Center for Theoretical Physics, Physics Department, Center for the
Study of Complex Systems, The University of Michigan, Ann Arbor, MI
48109-1040, USA}
\maketitle

\section{Introduction}

\begin{figure}[h]
\includegraphics[width=10cm]{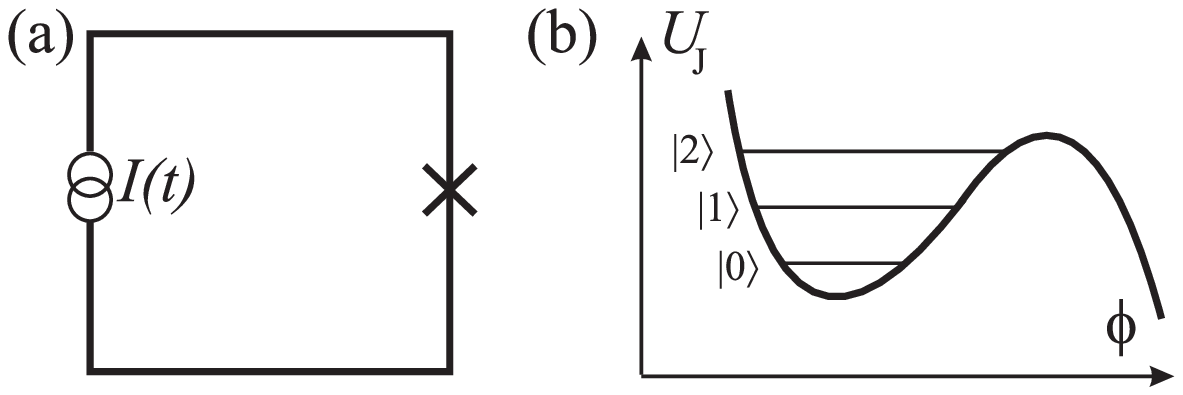}
\caption{Phase qubit (a) and its Josephson energy (b). The metastable states
$\left\vert 0\right\rangle $ and $\left\vert 1\right\rangle $ can be used as
qubit states.}
\label{scheme}
\end{figure}

Superconducting phase qubits \cite{You-Nori, Martinis02} provide a clear
demonstration of quantum coherent behaviour in macroscopic systems. They
also have a very simple design: a phase qubit is a current-biased Josephson
junction (see Fig. \ref{scheme}(a)), and its working states $|0\rangle $, $%
|1\rangle $ are the two lowest metastable energy levels $E_{0,1}$ in a local
minimum of the washboard potential. The transitions between these levels are
produced by applying an RF signal at a resonant frequency $\omega
_{10}=(E_{1}-E_{0})/\hbar \equiv \Delta E/\hbar $. The readout utilizes the
fact that the decay of a metastable state of the system produces an
observable reaction: a voltage spike in the junction or a flux change in a
coupled dc SQUID. In the three-level readout scheme (Fig. \ref{scheme}(b))
both $|0\rangle $ and $|1\rangle $ have negligible decay rates. A pulse at a
frequency $\omega _{21}=(E_{2}-E_{1})/\hbar $ transfers the probability
amplitude from the state $|1\rangle $ to the fast-decaying state $|2\rangle $%
. Its decay corresponds to a single-shot measurement of the qubit in state $%
|1\rangle $. Alternatively, instead of an RF readout pulse one can apply a
dc pulse, which increases the decay rate of $|1\rangle $.

One of the effects observed in driven phase qubits is Rabi oscillations \cite%
{Martinis02, Claudon04}: coherent transitions in a quantum two-level system
under the influence of a resonant perturbation, with a much lower frequency
dependence on the perturbation amplitude $A$ via $\Omega _{\mathrm{R}}=\sqrt{%
A^{2}+\delta ^{2}}$, where $\delta =\omega -\omega _{10}$ is the detuning of
the driving frequency $\omega $ from the resonance frequency $\omega _{10}$.
In resonance, $\Omega _{\mathrm{R}}=A$. Multiphoton Rabi oscillations, at $%
\omega _{10}=n\omega $ ($n$ stands for an integer), observed in such qubits
\cite{Wallraff03, Strauch07}, were also interpreted as a signature of
coherence.

The quantum coherent dynamics in phase qubits was tested by several
complementary methods (see, e.g., Refs. [%
\onlinecite{Martinis03,  Claudon04,
Strauch07,Lisenfeld07}]). In particular, in Ref. [\onlinecite{Martinis03}]
Rabi oscillations between the ground and excited state of the phase qubit
were measured by applying a 25 ns pulse at $\omega_{10}$ followed by a
measurement pulse at $\omega_{21}$. The probability to find the system in
the upper state oscillated with the amplitude of the resonant signal, as it
should in case of Rabi oscillations. In Ref. [\onlinecite{Lisenfeld07}] Rabi
oscillations were seen instead as the resonant pulse duration varied; a dc
readout was used.

However, it is known that the quantum mechanical behaviour of a quantum
two-level system can still be similar to the dynamical behaviour of the
classical non-linear oscillator \cite{Spreew, Zhu, Peano, Wei06, new, Alicki}%
. In particular, it was recently pointed out \cite{Gr-Jen-Cir, Marchese}
that due to the nonlinear behaviour of current-biased Josephson junctions, a
similar effect can arise in a purely classical way. Though direct tests -
\`{a} la Bell - of whether a given system is quantum or not are possible,
even their simplified versions (e.g., Ref. [\onlinecite{Alicki}]) are
demanding. This motivated us to further investigate the classical behaviour
of a phase qubit and find that there is a possibility to distinguish the
quantum Rabi oscillations from its classical double, by the shape of the
resonance, by the fact that the classical effect can be also produced by the
overtones, $n\omega _{10}$, of the resonance frequency, and by the sign of
the resonant frequency shift. (An observation that a non-Lorentzian shape of
the resonance should exclude a classical explanation was made already in
Ref. [\onlinecite{Berkley03}], where the spectroscopy of two coupled qubits
was performed. Also, a symmetric versus asymmetric Stark shift in a qubit
playing the role of a detector was proposed to distinguish the classical and
quantum behaviour of a nanomechanical oscillator \cite{Wei06}.) Classical
and quantum resonances, as a function of applied drives are also studied in
Ref. \cite{Savelev}.

\section{Model}

The phase qubit \cite{You-Nori, Martinis02} is a current-biased Josephson
junction. The Josephson potential, as a function of the phase difference $%
\phi $,%
\begin{equation}
U_{J}(\phi )=-\frac{\hbar I_{c}}{2e}(\cos \phi +\phi I_{\mathrm{dc}}/I_{c}),
\label{eq_U}
\end{equation}%
forms local minima, in which the quantized metastable levels serve as the
working states of the qubit. Here $I_{c}$ is the critical current of the
junction, and $I_{\mathrm{dc}}$ is the static bias current. A perturbation
can be produced by applying a time-dependent bias current, $I_{\mathrm{ac}%
}\sin \omega t$. In the quantum case, the system can be reduced to a
two-level model, described by the Hamiltonian \cite{Martinis03}%
\begin{equation}
\hat{H}=\frac{\Delta E}{2}\sigma _{z}+\frac{\hbar I_{\mathrm{ac}}\sin \omega
t}{\sqrt{2\Delta EC}}(\sigma _{x}+\chi \sigma _{z}),  \label{eq_quantum}
\end{equation}%
where $\sigma _{z,x}$ are Pauli matrices, $C$ is the capacitance of the
junction, and $\chi \approx 1/4$ in the relevant range of parameters. One
can obtain from here the Rabi oscillations (coherent oscillations of the
probability to find the system in the upper/lower state with the frequency $%
\Omega _{R}$) when excited near the resonance or at its subharmonics; the
shape of the resonance is a symmetric Lorentzian, as determined by, e.g.,
the average energy of the system versus the driving frequency $\omega $.

Unlike the flux qubit \cite{Mooij, OSZIN}, where the interlevel distance $%
\Delta E$ is determined by the tunneling, here $\Delta E$ is close to the
\textquotedblleft plasma\textquotedblright\ frequency $\omega
_{p}=[2eI_{c}/\hbar C]^{1/2}$ of small oscillations near the local minima of
the potential, Eq. (\ref{eq_U}): characteristically $\Delta E/\hbar \approx
0.95\omega _{p}$ \cite{Martinis02}. The same frequency determines the
resonance in the system in the classical regime. This means that, in
principle, the same AC signal could cause either the Rabi oscillations or
their classical double \cite{Gr-Jen-Cir}.

\section{Classical regime}

In the classical regime, the phase qubit can be described by the RSJJ
(resistively shunted Josephson junction) model \cite{Likharev, Barone}, in
which the equation of motion for the superconducting phase difference across
the junction, characterized by the normal (quasiparticle) resistance $R$,
reads
\begin{equation}
\frac{\hbar C}{2e}\frac{d^{2}\phi }{dt^{2}}+\frac{\hbar }{2eR}\frac{d\phi }{%
dt}+I_{c}\sin \phi =I_{\mathrm{dc}}+I_{\mathrm{ac}}\sin \omega t.
\end{equation}%
Introducing the dimensionless variables,
\begin{equation}
\tau =\omega _{p}t,\text{ \ }\gamma =\omega /\omega _{p},
\end{equation}%
we obtain
\begin{equation}
\ddot{\phi}+\alpha \dot{\phi}+\sin \phi =\eta +\epsilon \sin \gamma \tau ,
\label{eq_motion}
\end{equation}%
where
\begin{equation}
\alpha =\frac{\hbar \omega _{p}}{2eRI_{c}},\text{ }\eta =\frac{I_{\mathrm{dc}%
}}{I_{c}},\text{ }\epsilon =\frac{I_{\mathrm{ac}}}{I_{c}},
\end{equation}%
and the dot stands for the derivative with respect to $\tau $. The solution
is sought in the phase-locked Ansatz,
\begin{equation}
\phi (\tau )=\phi _{0}+\psi (\tau ),\text{ \ \ }\psi \ll 1.  \label{7}
\end{equation}%
We substitute Eq. (\ref{7}) in Eq. (\ref{eq_motion}) and expand $\sin \phi $
to third order, which yields
\begin{equation}
\ddot{\psi}+\alpha \dot{\psi}+\psi \cos \phi _{0}=\eta -\sin \phi _{0}+\frac{%
\sin \phi _{0}}{2}\psi ^{2}+\frac{\cos \phi _{0}}{6}\psi ^{3}+\epsilon \sin
\gamma \tau .  \label{eq_motion_2}
\end{equation}%
Therefore $\sin \phi _{0}=\eta $, and introducing
\begin{equation}
\gamma _{0}=[1-\eta ^{2}]^{1/4},
\end{equation}
we obtain
\begin{equation}
\ddot{\psi}+\alpha \dot{\psi}+\gamma _{0}^{2}\psi =\epsilon \sin \gamma \tau
+\frac{\eta }{2}\psi ^{2}+\frac{\gamma _{0}^{2}}{6}\psi ^{3},  \label{the_eq}
\end{equation}%
which describes the anharmonic driven oscillator [\onlinecite{LL}].

Here we briefly point out several features of the solution of Eq. (\ref%
{the_eq}) (for more details see Chap. 5 in Ref. [\onlinecite{LL}]).\newline

(i) The anharmonic driven oscillator, described by Eq. (\ref{the_eq}), \emph{%
is resonantly excited at any frequency }$\gamma \approx \frac{p}{q}\gamma
_{0}$, where $p$ and $q$ are integers. This however happens in higher
approximation in the driven amplitude $\epsilon $. At small amplitude $%
\epsilon $ the most pronounced resonances appear at $\gamma \approx \gamma
_{0}$ (main resonance), $\gamma \approx \gamma _{0}/2$ (anharmonic-type
resonance), and $\gamma \approx 2\gamma _{0}$ (parametric-type resonance).%
\newline

(ii) The amplitude $b$ of the small driven oscillations at the main
resonance is $b_{\max }=\epsilon /(\gamma _{0}\alpha )$. When this amplitude
is not small, then the phase-locked Ansatz becomes invalid. Then the
solution of Eq. (\ref{eq_motion}) describes the escape from the phase-locked
state, which means the appearance of a non-zero average voltage on the
contact. This voltage is proportional to the average derivative of the
phase, $\overline{\dot{\phi}}$.\newline

(iii) The position of the resonances for small oscillations is shifted due
to the anharmonicity of the potential. This, for the main resonance $\gamma
_{\mathrm{res}}$, is given by:
\begin{eqnarray}
\gamma _{\mathrm{res}}-\gamma _{0} &=&\varkappa b^{2}, \\
\varkappa &=&-\frac{3+2\eta ^{2}}{48\gamma _{0}^{3}}.
\end{eqnarray}%
Note that \emph{the resonance shift\ is negative} (e.g. as in [%
\onlinecite{Wei06}]).\newline

(iv) The shape of the resonances, as a function of the driven frequency $%
\gamma $, is essentially \emph{non-symmetrical}. The asymmetry of the main
resonance becomes pronounced at
\begin{equation}
\epsilon \gtrsim \frac{\gamma _{0}\alpha ^{3/2}}{\left\vert \varkappa
\right\vert ^{1/2}},
\end{equation}%
at small enough damping.\newline

(v) The\emph{\ parametric-type resonance} at $\gamma \approx 2\gamma _{0}$
appears when the damping is sufficiently low, i.e. at
\begin{equation}
\alpha <\frac{\eta \epsilon }{6\gamma _{0}^{3}}.
\end{equation}%
At the relevant parameters, $\eta \simeq 0.9$, $\gamma _{0}\simeq 0.6$, this
means the following:
\begin{equation}
\frac{\epsilon }{\gamma _{0}\alpha }>\frac{6\gamma _{0}^{2}}{\eta }\simeq 2,
\end{equation}%
which is fulfilled (see (ii)) when the solution close to the main resonance
corresponds to the escape from the phase-locked state.\newline

We note that for the anharmonic driven oscillator, described by Eq. (\ref%
{the_eq}), both the resonances at $\gamma _{0}/2$ and $2\gamma _{0}$ appear
due to the anharmonicity of the potential energy and are of the same order.
For the driven flux qubit in the classical regime \cite{OSZIN} the equation
for the phase variable $\theta (t)=\theta _{0}+\psi (t)$ can also be
expanded for small oscillations $\psi (t)$ about the value $\theta _{0}$;
restricting ourselves here to the linear in $\psi $ terms, the equation can
be rewritten in the form:
\begin{equation}
\ddot{\psi}+\alpha \dot{\psi}+\gamma _{0}^{2}[1-h\sin \gamma \tau ]\psi
=\epsilon \sin \gamma \tau .  \label{flux_qubit}
\end{equation}%
In this case the genuine parametric resonance at $\gamma \approx 2\gamma
_{0} $ appears due to the term containing $\sin \gamma \tau \cdot \psi $
\cite{LL}. This explains the prevailing of this resonance over the resonance
at $\gamma \approx \gamma _{0}/2$, due to the small anharmonicity of the
potential in Ref. [\onlinecite{OSZIN}].

\begin{figure}[h]
\includegraphics[width=14cm]{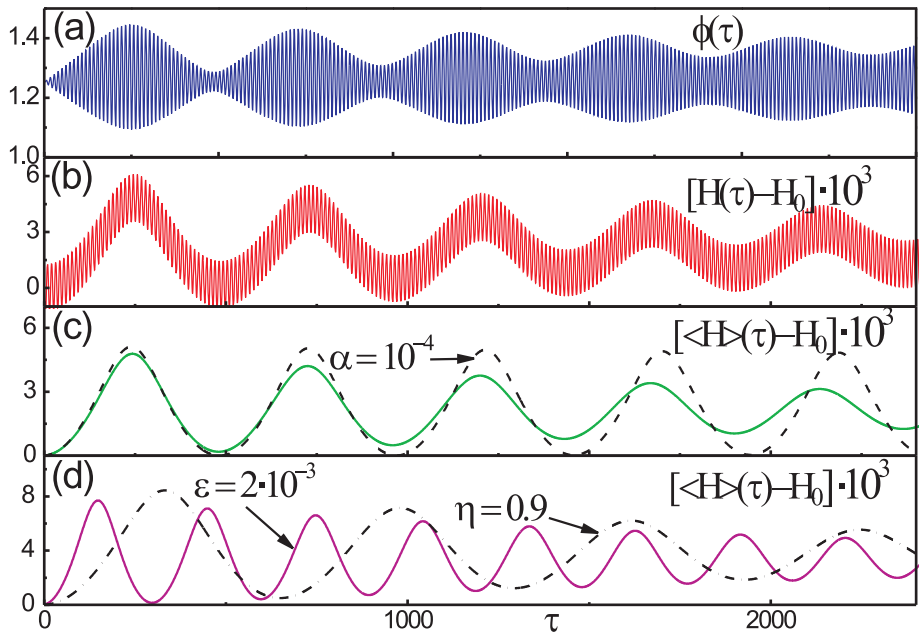}
\caption{(Color online). Rabi-type oscillations in current-biased Josephson
junctions: (a) and (b) show the time-dependence of the phase difference $%
\protect\phi $ and of the energy $H$, (c) presents the time-dependence of
the energy, averaged over the fast oscillations with period $2\protect\pi /%
\protect\omega $. All energies are shifted by their stationary value: $%
H_{0}=1-\protect\sqrt{1-\protect\eta ^{2}}-\protect\eta \arcsin \protect\eta
$. The parameters for the blue, red, and green curves are: $\protect\eta %
=0.95$, $\protect\alpha =10^{-3}$, $\protect\epsilon =10^{-3}$, and $\protect%
\gamma =\protect\gamma _{0}$; for the other curves in (c) and (d) only one
parameter was different from the above, for comparison. Namely: (c) dashed
black line $\protect\alpha =10^{-4}$, (d) solid violet line $\protect%
\epsilon =2\cdot 10^{-3}$, dash-dotted black line $\protect\eta =0.9$.}
\label{Figure1}
\end{figure}

Now we proceed to numerically solve the equation of motion (\ref{eq_motion})
for the relevant set of parameters close to the experimental case. We also
investigate the behaviour of the energy of the system \cite{Likharev, Barone}
\begin{equation}
H=\frac{1}{2}\dot{\phi}^{2}+1-\cos \phi -(\eta +\epsilon \sin \gamma \tau
)\phi ,
\end{equation}%
which determines the thermally activated escape probability from the local
minimum of the potential \cite{Gr-Jen-04} in Eq. (\ref{eq_U}). The classical
Rabi-like oscillations are displayed in Fig. \ref{Figure1}. In Fig. \ref%
{Figure1}(a) the modulated transient oscillations of the phase difference $%
\phi $ are plotted. These oscillations result in the oscillating behaviour
of the energy of the system as shown in Fig. \ref{Figure1}(b). Averaging
over fast oscillations, we plot in Fig. \ref{Figure1}(c) with green solid
curve the damped oscillations of the energy, analogous to the quantum Rabi
oscillations \cite{Gr-Jen-Cir}. These curves are plotted for the following
set of the parameters: $\eta =0.95$, $\alpha =10^{-3}$, $\epsilon =10^{-3}$,
and $\gamma =\gamma _{0}$. For comparison we also plotted the energy
averaged over the fast oscillations for different parameters, changing one
of these parameters and leaving the others the same. The dashed black curve
in Fig. \ref{Figure1}(c) is for the smaller damping, $\alpha =10^{-4}$; the
solid (violet) line and the dash-dotted (black) line in Fig. \ref{Figure1}%
(d) demonstrate the change in the frequency and the amplitude of the
oscillations respectively for $\epsilon =2\cdot 10^{-3}$ and $\eta =0.9$. We
notice that the effect analogous to the classical Rabi oscillations exist in
a wide range of parameters.

\begin{figure}[h]
\includegraphics[width=10cm]{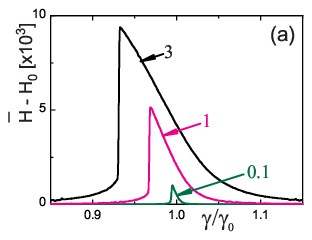} %
\includegraphics[width=10cm]{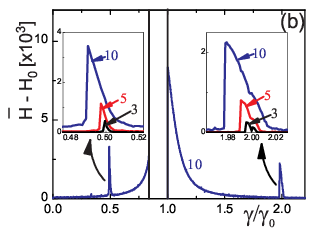}
\caption{(Color online). The time-averaged energy $\overline{H}-H_{0}$\
versus reduced frequency for relatively weak (a) and strong (b) driving.
Different values of the driving amplitude $\protect\epsilon $ (multiplied by
$10^{3}$) are shown by the numbers next to the curves. The parameters are: $%
\protect\eta =0.95$ and $\protect\alpha =10^{-4}$. In (b) the region between
the vertical black lines corresponds to the escape from the phase-locked
state.}
\label{Figure2}
\end{figure}

In Fig. \ref{Figure2} the effect of the driving current on the time-averaged
energy of the system is shown for different driving amplitudes: Fig. \ref%
{Figure2}(a) for weaker amplitudes, close to the main resonance, to show the
asymmetry and negative shift of the resonance; and Fig. \ref{Figure2}(b) for
stronger amplitudes, to show the resonances at $\gamma _{0}/2$ and $2\gamma
_{0}$ (which are also shown closer in the insets). We note that the
parametric-type resonance at $2\gamma _{0}$ originates from the third-order
terms when the solution of the equation for $\psi $ is sought by iterations
\cite{LL}; when there are two or more terms responsible for this resonance,
the respective resonance may become splitted, which is visible in Fig. \ref%
{Figure2}(b) for the lowest curve. An analogous tiny splitting of the
resonance was obtained for the driven flux qubit in Fig. 4 of Ref. [%
\onlinecite{OSZIN}].

\section{Quantum regime}

In the quantum regime, the phase qubit can be described by the Bloch
equations for the density matrix components. In order to take into account
the relaxation and dephasing processes, the corresponding rates $\Gamma _{%
\mathrm{relax}}\equiv \hbar \omega _{p}\cdot \lambda _{\mathrm{relax}}$\ and
$\Gamma _{\phi }\equiv \hbar \omega _{p}\cdot \lambda _{\phi }$\ are
included in the Liouville equation phenomenologically \cite{Blum}. Then the
evolution of the reduced density matrix $\widehat{\rho }$,\ taken in the
form
\begin{equation}
\widehat{\rho }=\frac{1}{2}\left[
\begin{array}{cc}
1+Z & X-iY \\
X+iY & 1-Z%
\end{array}%
\right] \text{,}
\end{equation}%
is described by the Bloch equations \cite{Blum, ShKOK}:
\begin{align}
\overset{\cdot }{X}& =-CY-\lambda _{\phi }X,  \label{eq1} \\
\overset{\cdot }{Y}& =-AZ+CX-\lambda _{\phi }Y,  \label{eq2} \\
\overset{\cdot }{Z}& =AY-\lambda _{\mathrm{relax}}\left( Z-Z_{0}\right) .
\label{eq3}
\end{align}%
where $A$ and $C$ stand for the off-diagonal and diagonal parts of the
dimensionless Hamiltonian:%
\begin{equation}
\frac{\hat{H}}{\hbar \omega _{p}}=\frac{\Delta E}{2\hbar \omega _{p}}\sigma
_{z}+\frac{\hbar I_{\mathrm{ac}}\sin \omega t}{\hbar \omega _{p}\sqrt{%
2\Delta EC}}(\sigma _{x}+\chi \sigma _{z})\equiv \frac{A}{2}\sigma _{x}+%
\frac{C}{2}\sigma _{z}.
\end{equation}%
From these equations we obtain $Z(\tau )$,\ which defines the occupation
probability of the upper level,
\begin{equation}
P(\tau )=\rho _{22}(\tau )=\frac{1}{2}(1-Z(\tau )).
\end{equation}%
We choose the initial condition to be $X(0)=Y(0)=0$, $Z(0)=1$, which
corresponds to the system being in the ground state; we also consider the
zero-temperature limit in which the equilibrium value of $Z$\ is $Z_{0}=1$.
\begin{figure}[h]
\includegraphics[width=11cm]{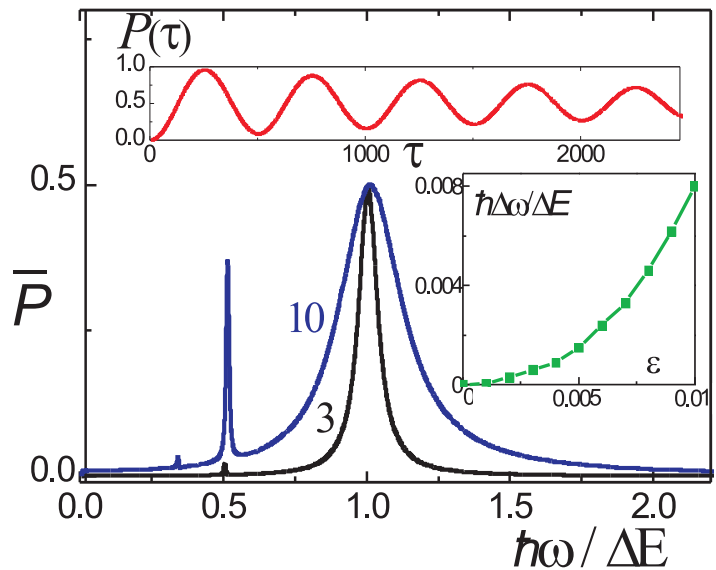}
\caption{(Color online). The time-averaged probability $\overline{P}$ of the
upper level to be occupied versus the driving frequency. The parameters used
here are: $\protect\eta =0.95$, $E_{J}/\hbar \protect\omega _{p}=300,$ $%
\Gamma _{\mathrm{relax}}/\hbar \protect\omega _{p}=\Gamma _{\protect\phi %
}/\hbar \protect\omega _{p}=3\cdot 10^{-4}$. Numbers next to the curves
stand for $\protect\epsilon $ multiplied by $10^{3}$. Upper inset: the time
dependence of the probability $P(\protect\tau )$. Right inset: the shift of
the principal resonance (at $\hbar \protect\omega \approx \Delta E$), where $%
\Delta \protect\omega =\protect\omega -\Delta E/\hbar $.}
\label{Figure3}
\end{figure}

When the system is driven close to resonance, $\omega \approx \Delta E$, the
upper level occupation probability $P(\tau )$\ exhibits Rabi oscillations.
The damped Rabi oscillations are demonstrated in the upper inset in Fig. \ref%
{Figure3}, which is analogous to the classical oscillations presented in
Fig. \ref{Figure1}. After averaging the time dependent probability, we plot
it versus frequency in Fig. \ref{Figure3} for two values of the amplitude,
demonstrating the multiphoton resonances. Figure \ref{Figure3} demonstrates
the following features of the multiphoton resonances in the quantum case:
(a) in contrast to the classical case, the resonances appear only at the
subharmonics, at $\hbar \omega \approx \Delta E/n$; (b) the resonances have
Lorentzian shapes (as opposed to the classical asymmetric resonances); (c)
with increasing the driving amplitude the resonances shift to the higher
frequencies -- the Bloch-Siegert shift, which has the opposite sign from its
classical counterpart. The Bloch-Siegert shift (the shift of the principal
resonance at $\hbar \omega \approx \Delta E$) is plotted numerically in the
right inset in Fig. \ref{Figure3}. Analogous shifts of the positions of the
resonances were recently observed experimentally \cite{Strauch07}.

\section{Conclusions}

In conclusion, the following criteria can be proposed to distinguish
classical Rabi-type oscillations in current-biased Josephson junctions: (1)
the appearance of resonances both at the overtones of the main resonant
frequency (e.g., $2\gamma _{0}$) and its subharmonic ($\gamma _{0}/2$); (2)
the asymmetric shape of the resonances; (3) a negative shift of the resonant
frequency when increasing the driving amplitude. In recent publications
these features were not reported: the multiphoton resonances were observed
\cite{Wallraff03, Strauch07}; the resonances observed have a Lorentzian shape \cite%
{Berkley03, Claudon04, comment} and have a positive frequency shift \cite%
{Strauch07}, which means that the observed resonant excitations in the
system were in the \emph{quantum} regime.

\begin{acknowledgments}
SNS acknowledges the financial support of INTAS under YS Fellowship Grant
No. 05-109-4479. ANO and SNS are grateful to Advanced Science Institute,
RIKEN, for hospitality. FN acknowledges partial support from the National
Security Agency (NSA), Laboratory Physical Science (LPS), Army Research
Office (ARO), and National Science Foundation (NSF) grant No. EIA-0130383.
FN and SS acknowledge partial support from JSPS-RFBR 06-02-91200, and
Core-to-Core (CTC) program supported by the Japan Society for Promotion of
Science (JSPS). SS acknowledges support from the Ministry of Science,
Culture and Sport of Japan via the Grant-in Aid for Young Scientists No.
18740224, the UK EPSRC via No. EP/D072581/1, EP/F005482/1, and ESF
network-programme \textquotedblleft Arrays of Quantum Dots and Josephson
Junctions\textquotedblright .
\end{acknowledgments}


\begin{thebibliography}{99}
\bibitem{You-Nori} You J Q and Nori F 2005 \textit{Physics Today} \textbf{58,%
} No. 11, 42

\bibitem{Martinis02} Martinis J, Nam S, Aumentado J and Urbina C 2002
\textit{Phys. Rev. Lett.} \textbf{89} 117901

\bibitem{Claudon04} Claudon J, Balestro F, Hekking F W J and Buisson O 2004
\textit{Phys. Rev. Lett.} \textbf{93} 187003

\bibitem{Wallraff03} Wallraff A, Duty T, Lukashenko A and Ustinov A V 2003
\textit{Phys. Rev. Lett.} \textbf{90} 037003

\bibitem{Strauch07} Strauch F W, Dutta S K, Paik H, Palomaki T A, Mitra K,
Cooper B K, Lewis R M, Anderson J R, Dragt A J, Lobb C J and Wellstood F C
2007 \textit{IEEE Trans. Appl. Supercond.} \textbf{17} 105

\bibitem{Martinis03} Martinis J M, Nam S, Aumentado J, Lang K M and Urbina C
2003 \textit{Phys. Rev. B} \textbf{67} 094510

\bibitem{Lisenfeld07} Lisenfeld J, Lukashenko A, Ansmann M, Martinis J M and
Ustinov A V 2007 \textit{Phys. Rev. Lett.} \textbf{99} 170504

\bibitem{Spreew} Spreeuw R J C, van Druten N J, Beijersbergen M W, Eliel E R
and Woerdman J P 1990 \textit{Phys. Rev. Lett.} \textbf{65} 2642

\bibitem{Zhu} Zhu Y, Gauthier D J, Morin S E, Wu Q, Carmichael H J and
Mossberg T W 1990 \textit{Phys. Rev. Lett.} \textbf{64} 2499

\bibitem{Peano} Peano V and Thorwart M 2004\textit{\ Phys. Rev. B} \textbf{70%
} 235401

\bibitem{Wei06} Wei L F, Liu Y X, Sun C P and Nori F 2006 \textit{Phys. Rev.
Lett.} \textbf{97} 237201

\bibitem{new} Rotoli G, Bauch T, Lindstrom T, Stornaiuolo D, Tafuri F and
Lombardi F 2007 \textit{Phys. Rev. B} \textbf{75} 144501

\bibitem{Alicki} Alicki R and Van Ryn N 2008 \textit{J. Phys. A} \textbf{41}
062001

\bibitem{Gr-Jen-Cir} Gr\o nbech-Jensen N and Cirillo M 2005\textit{\ Phys.
Rev. Lett.} \textbf{95} 067001

\bibitem{Marchese} Marchese J E, Cirillo M and Gr\o nbech-Jensen N 2007
\textit{Open Systems and Information Dynamics} \textbf{14} 189

\bibitem{Berkley03} Berkley A J, Xu H, Ramos R C, Gubrud M A, Strauch F W,
Johnson P R, Anderson J R, Dragt A J, Lobb C J and Wellstood F C 2003
\textit{Science} \textbf{300} 1548

\bibitem{Savelev} Savel'ev S, Hu X and Nori F 2006 \textit{New J. Phys. }%
\textbf{8} 105; Savel'ev S, Rakhmanov A L, Hu X, Kasumov A and Nori F 2007
\textit{Phys. Rev. B} \textbf{75} 165417

\bibitem{Mooij} Mooij J E, Orlando T P, Levitov L, Tian L, van der Wal C H
and Lloyd S 1999 \textit{Science }\textbf{285} 1036

\bibitem{OSZIN} Omelyanchouk A N, Shevchenko S N, Zagoskin A M, Il'ichev E
and Nori F \textit{arXiv}:0705.1768

\bibitem{Likharev} Likharev K 1986 \textit{Dynamics of Josephson Junctions
and Circuits} (New York: Gordon and Breach)

\bibitem{Barone} Barone A and Paterno G 1982 \textit{Physics and
Applications of the Josephson Effect} (New York: Wiley-Interscience)

\bibitem{LL} Landau L D and Lifshitz E M 1976 \textit{Mechanics} (Oxford:
Pergamon)

\bibitem{Gr-Jen-04} Gr\o nbech-Jensen N, Castellano M G, Chiarello F,
Cirillo M, Cosmelli C, Merlo V, Russo R and Torrioli G 2006 \textit{in
\textquotedblleft Quantum Computing in Solid State Systems\textquotedblright
}, eds. Ruggiero B, Delsing P, Granata C, Pashkin Y and Silvestrini P p.111
(Springer)

\bibitem{Blum} Blum K 1981 \textit{Density Matrix Theory and Applications}
(New York--London: Plenum Press)

\bibitem{ShKOK} Shevchenko S N, Kiyko A S, Omelyanchouk A N and Krech W 2005
\textit{Low Temp. Phys.} \textbf{31} 564

\bibitem{comment} We note passing by that in other types of qubits also the
Lorentzian-shaped multiphoton resonances were observed \cite{nak01, yaponci,
Oliver, multiphoton, Sillanpaa}

\bibitem{nak01} Nakamura Y, Pashkin Yu A and Tsai J S 2001 \textit{Phys.
Rev. Lett.} \textbf{87} 246601

\bibitem{yaponci} Saito S, Thorwart M, Tanaka H, Ueda M, Nakano H, Semba K
and Takayanagi H 2004 \textit{Phys. Rev. Lett.} \textbf{93} 037001

\bibitem{Oliver} Oliver W D, Yu Ya, Lee J C, Berggren K K, Levitov L S and
Orlando T P 2005 \textit{Science }\textbf{310} 1653

\bibitem{multiphoton} Shnyrkov V I, Wagner Th, Born D, Shevchenko S N, Krech
W, Omelyanchouk A N, Il'ichev E and Meyer H-G 2006\textit{\ Phys. Rev. B}
\textbf{73} 024506

\bibitem{Sillanpaa} Sillanp\"{a}\"{a} M, Lehtinen T, Paila A, Makhlin Yu and
Hakonen P 2007 \textit{J. Low Temp. Phys}. \textbf{146} 253; 2006 \textit{%
Phys. Rev. Lett.} \textbf{96} 187002
\end{thebibliography}
\end{document}